\documentclass{article}

\usepackage{graphicx}  % standard LaTeX graphics tool;
\usepackage{amsmath}   % For getting proper math eqns
\usepackage{amssymb}   % Used for getting various symbols eg. gtrsim, lesssim etc.

\def\eq#1{{Eq.~(\ref{#1})}}

\def\fig#1{{Fig.~\ref{#1}}}

  \title{The Physical Principle that determines the Value of the Cosmological Constant}
  \author{T. Padmanabhan\\
  IUCAA, Pune University Campus,\\
  Ganeshkhind, Pune- 411 007.\\
  {\small {email: paddy@iucaa.ernet.in}}
  }

  \date{ }  %% This command  will supress printing the date.   If date is required, comment out this line.

\begin{document}
  
  \maketitle
  
  \begin{abstract}
Observations indicate that the evolution of our universe can be divided into three epochs consisting of early time inflation, radiation (and matter) domination and the late time acceleration. One can associate with each of these epochs a number $N$ which is the phase space volume of the modes which cross the Hubble radius during the corresponding epoch. This number turns out to be (approximately) the same for the cosmologically relevant ranges of the three epochs.  When the initial de Sitter space is characterized by the Planck length, the natural value for $N$ is $4\pi$. This allows us to determine the cosmological constant $\Lambda$, which drives the late time acceleration, to be $\Lambda L_P^2 = 3 \exp(-24\pi^2 \mu )$ where $\mu$ is a number of order unity. This expression leads to the observed value of cosmological constant for $\mu \approx 1.19$. The implications are discussed.
 \end{abstract}

 %%  Start the sections
 %\section{ }
 
 The current observations indicate that the radiation (and matter) dominated epoch of the universe is sandwiched between two asymptotic de Sitter epochs of expansion, usually identified with the inflationary era and the epoch of late time acceleration. 
These two de Sitter phases are characterized by two length scales $L_{\rm UV}$ and $L_{\rm IR}$ corresponding to the respective Hubble radii.
 Given the fact that de Sitter geometry is invariant under time translation, we can interpret the radiation (and matter) dominated phase of the universe as a transition state which connects the two ``equilibrium'' (steady) states of the geometry. In such a picture, it is natural to set the length scale for inflationary phase $L_{\rm UV}$ to be the Planck length $L_P$ and the length scale of the accelerating phase $L_{\rm IR}$ to be $(3/\Lambda)^{1/2}$ where $\Lambda $ is the cosmological constant. The ratio of these two length scales is conveniently expressed in terms of the dimensionless number
\begin{equation}
 \Lambda L_P^2  \approx 3\times 10^{-122}\approx 3\times e^{-281}
\end{equation}
where the numerical value is determined by observations (with $\Omega_{\rm DE} \approx 0.7$ and $h\approx 0.7$). 

An important question in theoretical physics is to determine this numerical value from first principles \cite{ccprob}. If a fundamental principle can be found which allows the determination of this number, then all the conventional difficulties associated with the cosmological constant (e.g., why is it fine-tuned, why does it dominate the universe now, etc. etc.) will vanish. I will describe a principle which allows us to express this number in the form
\begin{equation}
 \Lambda L_P^2 = 3 \exp(-24\pi^2 \mu )
\label{th1}
\end{equation}
where $\mu$ is a number of order unity and, in principle, calculable. \textit{Observations match with the above expression when $\mu \approx 1.19$ which I consider to be extremely encouraging.} Even for the ``natural'' choice of $\mu =1$, \eq{th1} predicts 
$\log(\Lambda L_P^2/3) = - 103$  compared to the observed value of $-122$. I do not know of any other attempt which could express the number $\Lambda L_P^2$ essentially in terms of $e,\pi,...$ etc. and get this close the observed value!

\begin{figure}
 \begin{center}
  \includegraphics[scale=0.5]{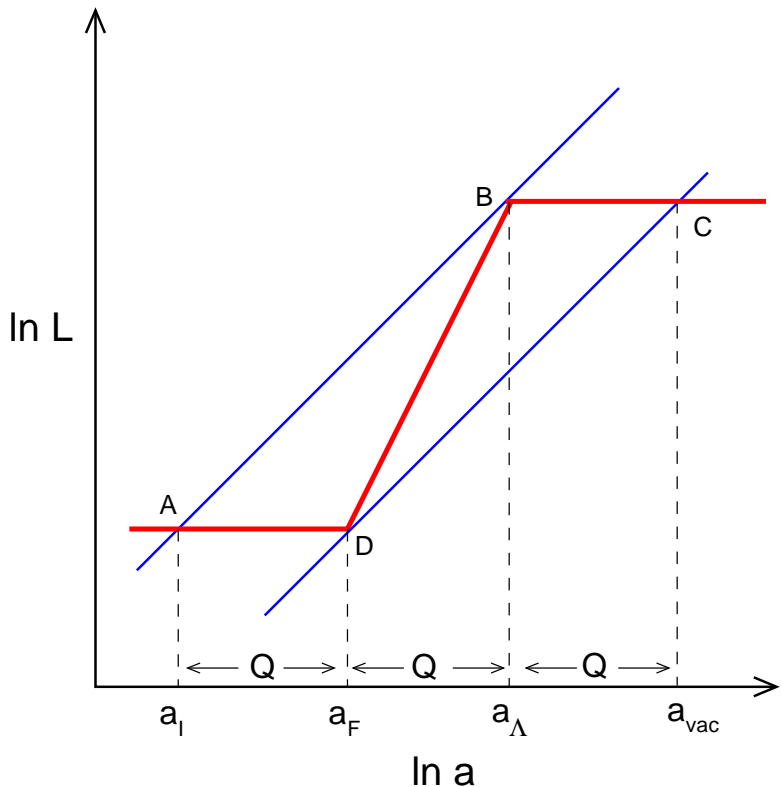}
 \end{center}
\caption{The relevant length scales in a universe characterized by a radiation dominated epoch sandwiched between two de Sitter phases. See text for details.}
\label{fig:detcc1}
\end{figure}

Let me now describe how \eq{th1} is obtained. I have shown in \fig{fig:detcc1}  the relevant length scales involved with our universe. The thick red line ADCB denotes the Hubble radius $H^{-1}(a)=(\dot a/a)^{-1}$ which is constant asymptotically during the early inflationary epoch ($a<a_F$) and the late time accelerating phase ($a>a_\Lambda$). In the intermediate epoch ($a_F<a<a_\Lambda$), the Hubble radius DB grows as $a^2$ if the universe is radiation dominated. (To be precise, there is a regime close to B when the universe becomes matter dominated which I have ignored for the moment and will comment on it later on. Note that this phase lasts only for about 4 decades while the radiation dominated phase lasts for about $24$ decades.) 

In principle, one can extend the asymptotic de Sitter phases (which represent the universe in a steady, time translation invariant state) into the past and future as long as one wants. However, there are two natural boundaries to these de Sitter phases (for a detailed discussion, see Ref.~\cite{tpbj}). The boundary A (at $a=a_I$) in the inflationary phase is determined as follows. We know that cosmologically relevant modes exit the Hubble radius during the inflationary phase and then re-enter the Hubble radius during the radiation (and matter) dominated phase. (The proper wavelengths of these modes grow linearly with $a$ and will be lines of unit slope in \fig{fig:detcc1}.)  \textit{All} the modes which exit the Hubble radius during the inflationary phase will re-enter the Hubble radius later on if there is  \textit{no} late time accelerating phase for the universe. But in the presence of late time acceleration, there is a mode with a critical wavelength (shown by the line AB, increasing linearly with $a$) which just skirts entering the Hubble radius. Given the  points D and B, the point A is determined by drawing a unit slope line through B.  Elementary geometry tells us that the universe expands by the same factor $Q\equiv (a_F/a_I)=(a_\Lambda/a_F)$ during AD and DB.  

The natural boundary C (at $a=a_{\rm vac}$) in the late time acceleration phase  is determined by a different consideration. We know that the de Sitter phase is associated \cite{GH} with the Gibbons-Hawking temperature $T_{\rm dS}= (H/2\pi)$. As the universe expands, the CMBR temperature will keep decreasing $T_{\rm cmb}(a) \propto (1/a)$ and will eventually fall below the de Sitter temperature $T_{\rm dS}$ after which the temperature of the universe will be essentially dominated by the de Sitter vacuum noise. This determines the point C through the condition $T_{\rm cmb}(a_{\rm vac})=T_{\rm dS}$. 

If we now take the initial de Sitter phase to be characterized by the Planck length (i.e., $H^{-1} = L_P$), then it is natural to assume that the reheating temperature at the end of Planck scale inflation is given by the  de Sitter temperature $T_P = 1/(2\pi L_P)$. In that case it is again elementary to show that the unit slope line \cite{comment1} drawn from C passes through D. So the relevant part of late time acceleration also lasts for an expansion by factor $Q=(a_{\rm vac}/a_\Lambda)$. 

In such a simple scenario, all the relevant physics is contained within the \textit{cosmic parallelogram} ADCB with the universe expansing by the \textit{same} factor $Q$ during each of the three epochs. Given the fact that $H^{-1}\propto a^2$ during the radiation dominated phase DB, we can relate the Hubble radius $H_\Lambda^{-1}=(3/\Lambda)^{1/2}$ at B with the Hubble radius $H_P^{-1}=L_P$ at D by $(H_P/H_\Lambda)=(a_\Lambda/a_F)^2=Q^2$. That is,  $(H_\Lambda/H_P)^2=(1/3)(\Lambda L_P^2)=Q^{-4}$ which allows us to relate the numerical value of the cosmological constant $\Lambda$ to $Q$ by:
\begin{equation}
\Lambda L_P^2=3Q^{-4}
\label{ccQ}
\end{equation} 
I will now describe a physical principle to determine $Q$.

To do this let us consider the modes (specified by co-moving wave vectors $k$) which exit the Hubble radius during AD. They  will enter the Hubble radius during DB and will again exit the Hubble radius during BC. \textit{The number of modes inside the Hubble volume which does this is a characteristic number for our universe.} Let us  calculate the total number of modes which cross the Hubble radius in the time interval $(t_1,t_2)$ or, more conveniently, when the expansion factor is in the range $(a_1,a_2)$. Since the phase space density of number of modes in the \textit{co-moving } Hubble  volume $V_{com}=4\pi/3H^3a^3$ is given by the integral of $dN=V_{com}d^3k/(2\pi)^3=V_{com}k^3/(2\pi^2)d\ln k$,
we need to compute the integral over the relevant range of $k$. (We, of course, get the same expression if we use the proper volume and proper wave number instead of co-moving variables). If we take the condition for horizon crossing to be $k=Ha$ then we get:
\begin{equation}
N(a_1,a_2)\equiv\int \frac{d^3xd^3k}{(2\pi)^3}= \int\frac{V_{com}d^3k}{(2\pi)^3}=\frac{2}{3\pi}\ln (Ha)\Big|_{a_1}^{a_2}
\label{N12}
\end{equation}  
where we have used $V_{com}=4\pi/3H^3a^3$ and $k=Ha$. We will choose $a_1<a_2$ or  $a_1>a_2$ depending on whether the  mode is exiting the Hubble radius or entering the Hubble radius so as to  make $N(a_1,a_2)$ positive. (Note that, during exit, $k_1^{-1}>k_2^{-1}$, implies $a_1<a_2$ while during the entry $k_1^{-1}>k_2^{-1}$ implies $a_1>a_2$.) 

The  relation in \eq{N12} is valid in general but it leads to something interesting during de Sitter and radiation dominated phases: In the de Sitter phase with constant $H$ we have $Ha\propto a$ while 
in the radiation dominated phase $H\propto a^{-2}$, so that $Ha\propto 1/a$. Therefore, during each of the three phases of the universe shown in Fig. 1, 
the total number of modes which cross the Hubble radius remains a constant:
\begin{equation}
 N(a_I,a_F) = N(a_\Lambda,a_F)= N(a_\Lambda,a_{\rm vac})\equiv N
\label{threeN}
\end{equation} 
given by:
\begin{equation}
 N=\frac{2}{3\pi}\ln Q
\label{basicN}
\end{equation}
 So the equality of ratios of expansion factors in the three phases translates to the equality of   the number of modes $N$ in a Hubble volume which crosses the Hubble radius in the three phases. \textit{This number $N$ is a characteristic, (approximately) conserved, quantity for the universe during the three phases.}

The above relation was obtained by assuming that there is an abrupt change of slope of the Hubble radius at D and B and ignoring the matter domination phase. Correcting for matter domination and including the the smooth transition at B is algebraically trivial since we know the behavior of our universe around B. The transition at the end stage of a Planck scale inflation with re-heating, emergence of matter etc. at D is more uncertain. We have also assumed that the condition for Hubble radius crossing is $k=Ha$. This is equivalent to taking the co-moving length scale corresponding to $k$ to be $1/k$; one could have taken this to be $2\pi/k, \pi/k, ...$ etc. with the same level of uncertainty. All these effects 
 could introduce order unity corrections to the expression for $N$ and to remind ourselves of this fact we will rewrite the expression in \eq{basicN} as
\begin{equation}
 \mu N=\frac{2}{3\pi}\ln Q
\label{basicN1}
\end{equation}
where $\mu$ is expected to be number of order unity. Substituting in \eq{ccQ} we can relate the value of the cosmological constant to $N$ by:
\begin{equation}
\Lambda L_P^2=3Q^{-4}=3\exp(-6\pi\mu N)
\label{ccfinal} 
\end{equation} 
\textit{I consider this an important result in its own right.} It reduces the problem of understanding the numerical value of cosmological constant to a more manageable problem of understanding a particular value for $\mu N$ for our universe. 

The natural value for $N$ is just $4\pi$ in our scenario. To see this, note that during the Planck scale inflation, the surface area of the Hubble sphere is $4\pi L_P^2$ and it is reasonable to assume that the total number of modes crossing this Hubble radius during the Planck scale domain should be of the order of $4\pi L_P^2/L_P^2=4\pi$. (There could again be an order unity factor which we will absorb into $\mu$. I will say more about this later on.) Using $N=4\pi$ in \eq{ccfinal}, we find:
\begin{equation}
\Lambda L_P^2=3\exp(-24\pi^2\mu)
\label{ccnumb} 
\end{equation}
As I said at the beginning, even for $\mu=1$ this gives $\Lambda L_P^2=3\times 10^{-103}$ which is within striking distance of the observations and far better than what any other model has achieved. The smallness of the cosmological constant is now related to its exponential dependence on $N$ plus the fact that $24\pi^2$ is a rather large number! One can get the correct, observed, value of the cosmological constant for $\mu=1.19$, which, as advertised, is an order unity number:
\begin{equation}
\Lambda_{obs} L_P^2=3\exp(-24\pi^2\mu) \qquad (\mu=1.19)
\label{ccnumb1} 
\end{equation}
This is the second result of this paper. I will now make several comments about the results and embed them in a broader context.

Let me first dispose of some technical points. Usually, one does not consider the Planck scale inflationary scenario because of the claims in the literature that it produces too much of gravitational wave perturbations. What is actually provable is that, if one considers spin-2 perturbations\textit{ within the framework of normal continuum field theory} in an inflationary background, then the primordial gravitational waves generated will violate the observational bound if  the inflation scale is close to the Planck scale. But this is not a convincing argument because, as we go close to the Planck scale, one cannot trust \textit{continuum field theory} of the spin-2 field and the results based on it. In fact, there are suggestions in the literature \cite{TPseshTP} that this problem goes away if one considers corrections to  propagators arising from quantum gravitational effects in the form of a cut-off at the Planck scale. While these are just toy models, it prevents me from taking the gravitational wave bounds seriously to exclude Planck scale inflation.

Second, the computation of $N$ in \eq{N12} only used one (e.g. scalar) degree of freedom and one might think that we should  multiply it by the effective number of species, $g_{\rm eff}$, at Planck scale. We do not know what this number is but it turns out to be irrelevant in the picture I have in mind. I consider the transition at D to be the emergence of space along with the emergence of matter degrees of freedom (which leads to the radiation dominated era) from some other pre-geometric degrees of freedom. I then expect the equipartition of gravitational and matter degrees of freedom to set the \textit{total} matter degrees of freedom $g_{\rm eff} N$ to some specific value like $4\pi$. So, the factor $g_{\rm eff}$ does not play any role in the final expression for $\Lambda L_P^2$. (That is, it will modify the intermediate equations \eq{basicN}, \eq{basicN1} and \eq{ccfinal} by changing $N$ to $ g_{\rm eff}N$ but the final result in \eq{ccnumb} will not change when we set $g_{\rm eff}N = 4\pi$.) In fact, the initial de Sitter phase, with the energy scale equal to the Planck energy, needs quantum gravitational inputs for its proper treatment and cannot be considered as the usual inflation driven by a scalar field, etc.

Third, there is a curious relation between $N$ and the entropy one can associate with modes that cross the Hubble radius. If $dN$ is the number of modes which cross the Hubble radius during the period when expansion factor changes by $da$, then they contain the energy $dE=(k/a)dN$ if we treat the modes as (effectively) massless. If we associate a temperature $T=H/2\pi$ with the Hubble radius, then we can associate an entropy $dS=dE/T=2\pi(k/Ha)dN=2\pi dN$ with these modes. So the number of modes $N(a_1,a_2)$ which cross the Hubble radius during $a_1<a<a_2$ is related to an entropy $S(a_1,a_2)=2\pi N(a_1,a_2)$. In terms of $S$, the \eq{basicN} becomes $Q=\exp(3\pi N/2)=\exp(3S/4)$. Equivalently,
\begin{equation}
e^S=Q^{\frac{4}{3}}=\left(\frac{a_2}{a_1}\right)^{\frac{4}{3}}
\end{equation} 
which relates the expansion factor to this entropy. It is possible to reformulate our analysis in terms of $S$ which I hope to address in a separate publication. 

Fourth, once the basic idea is grasped, it can be presented in many different perspectives. For example, one can \textit{start with} a radiation dominated epoch during which the universe expands by a factor $Q$ and express it in terms of the number $N$ of modes inside the Hubble volume which cross the Hubble radius during this epoch, obtaining $Q=\exp(3\pi \mu N/2)$. If we assume that these modes exit the Hubble radius during the earlier inflationary era and later accelerating phase, then we obtain the cosmic parallelogram ADCB relevant to these modes. We also obtain the relation $\Lambda L_P^2=3Q^{-4}=3\exp(-6\pi\mu N)$ determining the ratio of two de Sitter length scales in terms of $N$. Different universes with these properties are then characterized by different values for $N\mu$. 
 
 Fifth, I do not insist that $N$ be an integer because it is computed using a continuum field theory in \eq{N12}. In a more rigorous calculation, one would like to study field modes inside a Hubble sphere with specific boundary conditions and compute this number. We do not know how to do this correctly since we do not understand how some pre-geometric variables lead to emergent degrees of freedom at Planck scales. I hope to discuss this and the implications for CMBR anisotropy at horizon scales due to the change in the boundary conditions in a future work.
 
 Finally, as mentioned earlier, there are other (trivially calculable) order unity effects which were absorbed into $\mu$. These include: (i) The ambiguity in the condition for crossing the Hubble radius (taken to be $k= Ha$ while one could have used $(k/2\pi) , \ (k/\pi), ....$ etc.); (ii) the effect of the matter dominated phase; (iii) the effect due to smoothening the sharp transition which was assumed at B. I have not bothered to exhibit these results in gory detail since we have a much larger uncertainty in modeling the transition at D  from first principles. 

Let us now turn to the conceptual aspects of this approach. I consider the above analysis as a \textit{program capable of determining the numerical value of $\Lambda L_P^2$}. Such a program has the following ingredients: 
\begin{itemize}

 \item The universe is described by two fundamental length scales  $L_P$ and $\Lambda^{-1/2}$ or --- equivalently --- by one length scale $L_P$ and the dimensionless ratio $\Lambda L_P^2$. This ratio needs to be determined by a physical principle and the fact that it is very small should become obvious when this principle is implemented properly.

\item Time translation invariance of the geometry suggests that de Sitter spacetime qualifies as some kind of ``equilibrium''
configuration. Given the two length scales, one can envisage two de Sitter phases for the universe, one governed by $H=L_P^{-1}$ and the other governed by $H=(\Lambda/3)^{1/2}$. Of these, I would expect the Planck scale inflationary phase to be an unstable equilibrium causing the universe to make a transition towards the second de Sitter phase governed by the cosmological constant. The transient stage is  populated by matter emerging along with classical geometry around the point D
in \fig{fig:detcc1}.

\item Such a cosmological model is characterized by a number $N$ related to $\Lambda L_P^2$ by \eq{ccfinal}. This $N$ has a direct physical interpretation as the number of modes within a Hubble volume which cross the Hubble radius during any of the three phases of evolution of the universe.  \textit{Because $N$ has a direct physical meaning}, this translates the problem of determining a very small number $\Lambda L_P^2$ to the problem of determining a more manageable number $(1/6\pi) \ln (\Lambda L_P^2/3)$ which is of order 10 for our universe. 

\item I have given an argument as to why $N$ is of order $4\pi$. This is probably the weakest part of the paper; but it necessarily has to be so since I have not solved the problem of quantum gravity or how matter and space emerges from some pre-geometric variables. But even without such a detailed knowledge one can argue that the numerical value cannot be far widely off from the result $N = 4\pi$.  Given a better model for quantum gravity, one should be able to calculate $\mu$ and obtain a more precise numerical value for $\Lambda$.

\end{itemize}

This program resonates well with the idea that gravity is an emergent phenomenon. There is considerable evidence in classical gravitational theories (for a review, see Ref.~\cite{TPreview}) that the \textit{field equations of gravity} have the same status as the equations describing emergent phenomena like elasticity or fluid mechanics. A more ambitious task will be to think of \textit{spacetime itself} as emergent rather than just the field equations. I have given arguments elsewhere \cite{tpholocosmo1} as to why this is possible in (but \textit{only} in) the context of cosmology in which there is a preferential choice of the time coordinate and the Lorentz frame allowing the luxury of treating \textit{space} as emergent as cosmic time evolves. This, in turn, is related to the fact that, purely experimentally, our universe exhibits much larger symmetry (e.g., Lorentz invariance, general covariance) at smaller scales while singling out a specific Lorentz frame at very large scales. If we think of cosmic expansion as the acausal emergence of space at the Hubble scale, then one could imagine some kind of thermodynamic equipartition of pre-geometric variables having already taken place at small scales (resulting in higher level of symmetry, emergent description of Einstein's equations, etc.) while such an equipartition has not yet occurred at the largest scales. This approach considers the cosmological description as fundamental rather than as a specific solution to the gravitational field equations. In fact, the equation governing the expansion of the universe can be written \cite{tpholocosmo1}, from first principles, in the form 
\begin{equation}
\frac{dV}{dt} = L_P^2 (N_{\rm sur} - \epsilon N_{\rm bulk}) \qquad \epsilon=\pm 1
\end{equation}
where $V$ is the volume of the Hubble sphere, $T= H/2\pi$ is the related temperature and
\begin{equation}
N_{\rm sur} = 4\pi \frac{H^{-2}}{L_P^2};\qquad 
N_{\rm bulk} =-\epsilon\frac{E}{(1/2)k_BT}
\end{equation}
are the degrees of freedom associated with the surface and the bulk.
This equation shows that the cosmic expansion is driven by the ``holographic discrepancy'' $(N_{\rm sur} - \epsilon N_{\rm bulk})$ between the surface and bulk degrees of freedom. When this discrepancy vanishes, one obtains the de Sitter solution.
(This idea is described in detail in Ref.~\cite{tpholocosmo1} and in the first paper in Ref.~\cite{tpbj}.)
In such a picture, matter and geometrical degrees of freedom must emerge simultaneously from the pre-geometric variables and one would indeed \textit{expect} a relationship between $g_{\rm eff} N$ and quantum gravitational physics.  I  consider my choice $g_{\rm eff} N = 4\pi$ to capture this (as yet unknown) part of physics.

The acceptance of the arguments in this paper provides a route for resolving what is often considered to be a major challenge in theoretical physics. I believe that the final solution to the cosmological constant problem will \textit{only} require refining the various ingredients described in the itemized list above. In particular, we need to accept the existence of two length scales in our universe and look for a first principle argument to determine the ratio between these two scales.

I thank Sunu Engineer for useful discussions. My research is partially supported by the J.C.Bose research grant of DST, India. 

 %% References begins


\begin{thebibliography}{000}

\bibitem{ccprob} For an excellent theoretical review, see J. Martin, [arXiv:1205.3365]; for a classification of approaches to cosmological constant problem, see e.g. S. Nobbenhuis, [arXiv:gr-qc/0411093].

\bibitem{tpbj} T. Padmanabhan, \textit{Res. Astro. Astrophys.}, \textbf{12}, 891 (2012) [arXiv:1207.0505];
                               \textit{Gen.Rel.Grav.,} \textbf{40}, 529-564 (2008) [arXiv:0705.2533]; 
 J. D. Bjorken, (2004),  [arXiv:astro-ph/0404233].

\bibitem{GH} G. W. Gibbons, S. W. Hawking, \textit{Phys Rev. }, \textbf{D 15}, 2738 (1977);
D. Lohia,  \textit{J. Phys.},  \textbf{ A 11}, 1335 (1978).


\bibitem{comment1} As an aside, note that I do \textit{not} determine C by drawing a unit slope line from D. Instead, the point C is determined by the condition $T_{\rm dS} = T_{\rm cmb}$. A unit slope line drawn from C will not, in general, pass through the point D which is \textit{independently} specified as  the end point of inflation. It passes through D in this specific scenario because we have ignored the matter dominated phase and taken the reheating temperature to be set by the Planck scale. But even in a more realistic scenario, the three phases of expansion last for approximately equal number of decades; see, for details,  Ref.~\cite{tpbj}.

\bibitem{TPseshTP} T.Padmanabhan, \textit{Phys. Rev. Letts.}, \textbf{60}, 2229 (1988);
T.Padmanabhan, T.R. Seshadri and T.P. Singh,  \textit{Phys. Rev.} \textbf{D 39}, 2100 (1989). 

\bibitem{TPreview} T.Padmanabhan,\textit{ AIP Conf. Proc.}, \textbf{1483}, 212 (2012) [arXiv:1208.1375];
                                 \textit{Rep. Prog. Phys.}, \textbf{73}, 046901  [arXiv:0911.5004].

\bibitem{tpholocosmo1}T. Padmanabhan, \textit{Emergence and Expansion of Cosmic Space as due to the Quest for Holographic Equipartition }[arXiv:1206.4916].


 \end{thebibliography}
 \end{document}